\documentclass[acmtiis,screen,sigconf]{acmart}

\acmISBN{}
\acmBooktitle{} 
\acmConference{}{}
\acmArticle{}

\renewcommand\footnotetextcopyrightpermission[1]{} %
\AtBeginDocument{%
  }

\acmISBN{978-1-4503-XXXX-X/18/06}

\usepackage{subcaption}
\usepackage{_macros}
\usepackage{multirow}
\usepackage{array,booktabs}
\usepackage{tabu}
\usepackage{tabularray}
\usepackage{nicematrix}
\usepackage[symbol]{footmisc}

\definecolor{BLACK}{rgb}{0, 0, 0}

\begin{document}

\title{Panda or not Panda? Understanding Adversarial Attacks with Interactive Visualization}

\author{Yuzhe You}
\orcid{0009-0004-7830-4239}
\affiliation{
  \institution{University of Waterloo}
  \streetaddress{200 University Ave W}
  \city{Waterloo}
  \state{Ontario}
  \country{Canada}
}
\email{y28you@uwaterloo.ca}

\author{Jarvis Tse}
\orcid{0000-0003-4001-4610}
\affiliation{%
  \institution{University of Waterloo}
  \streetaddress{200 University Ave W}
  \city{Waterloo}
  \state{Ontario}
  \country{Canada}
}
\email{j47xie@uwaterloo.ca} 

\author{Jian Zhao}
\orcid{0000-0001-5008-4319}
\affiliation{%
  \institution{University of Waterloo}
  \streetaddress{200 University Ave W}
  \city{Waterloo}
  \state{Ontario}
  \country{Canada}
}
\email{jianzhao@uwaterloo.ca}

\renewcommand{\shortauthors}{You et al.}

\begin{abstract}
  Adversarial machine learning (AML) studies attacks that can fool machine learning algorithms into generating incorrect outcomes as well as the defenses against worst-case attacks to strengthen model robustness. 
  Specifically for image classification, it is challenging to understand adversarial attacks due to their use of subtle perturbations that are not human-interpretable, as well as the variability of attack impacts influenced by diverse methodologies, instance differences, and model architectures. 
  Through a design study with AML learners and teachers, we introduce \name{}, a multi-level interactive visualization system that comprehensively presents the properties and impacts of evasion attacks on different image classifiers for novice AML learners. 
  We quantitatively and qualitatively assessed \name{} in a two-part evaluation including user studies and expert interviews.
  Our results show that \name{} is not only highly effective as a visualization tool for understanding AML mechanisms, but also provides an engaging and enjoyable learning experience, thus demonstrating its overall benefits for AML learners.
\end{abstract}



\renewcommand{\sd}[1]{{\small $\sigma=#1$}}
\newcommand{\ttest}[2]{{\small $t=#1,p=#2$}}
\newcommand{\name}{\textsc{AdvEx}}
\newcommand{\re}[1]{\textcolor{BLACK}{#1}}
\newcommand{\ree}[1]{\textcolor{BLACK}{#1}}

\maketitle

\begin{figure*}[!t]
\centering
  \includegraphics[width=\linewidth]{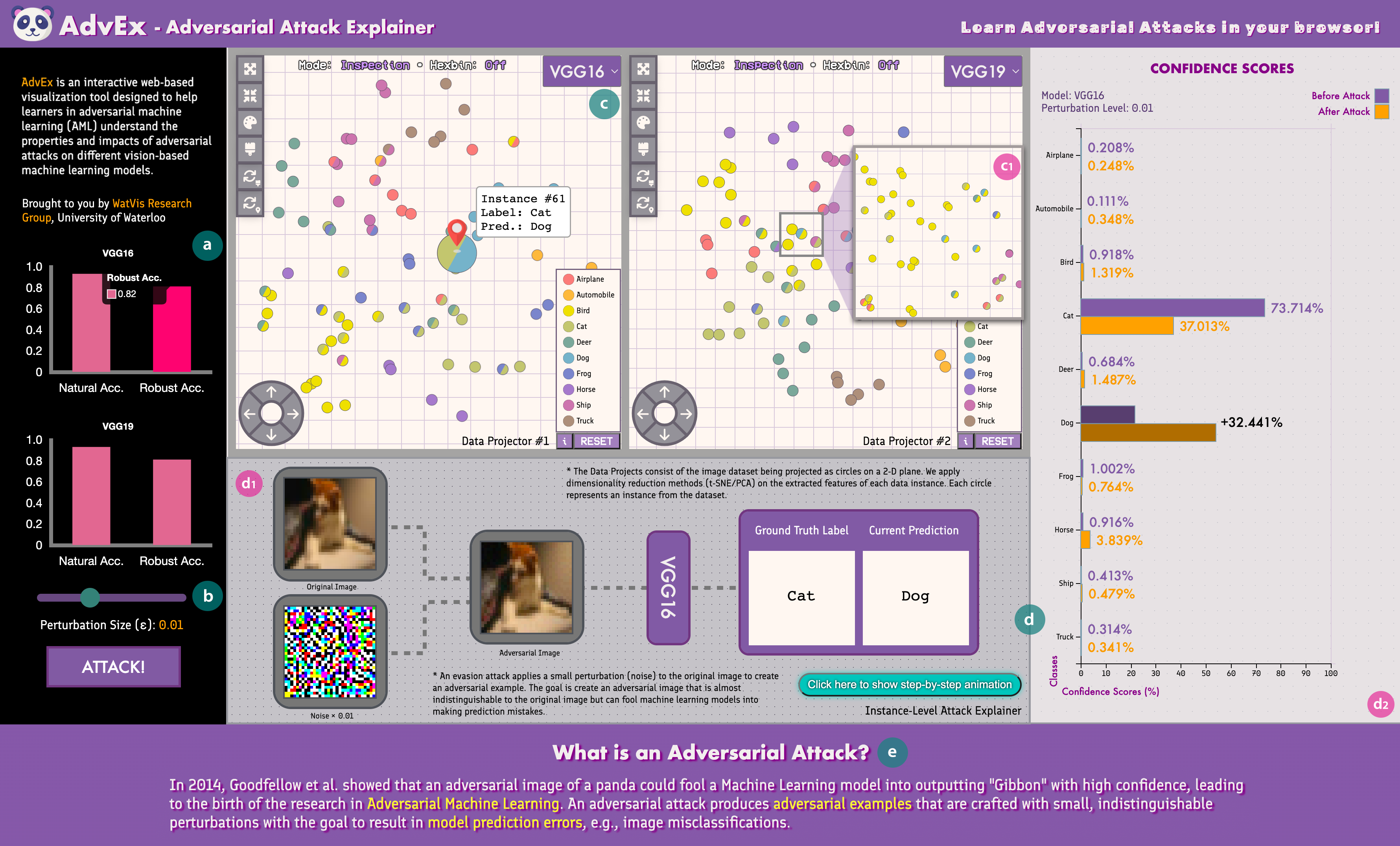}
  \vspace{-6mm}
  \caption{\name{} user interface: (a) Robustness Analyzers that display the models' prediction accuracy pre- and post-attack; (b) Perturbation Adjuster that initiates the attack sequence with specified magnitude; (c) Data Projectors that visualize data embeddings in a 2-D latent space; (d) Instance-level Attack Explainer that displays in-depth information of the highlighted instance; (e) General Information Provider that provides more background on \name{} and AML.}
  \label{fig:teaser}
  \vspace{-5mm}
\end{figure*}

\vspace{-3mm}
\section{Introduction} \label{sec:intro}

Adversarial \emph{evasion attacks} produce deceptive inputs (\eg, adversarial images) that are subtly altered with human-imperceptible perturbations to fool machine learning (ML) models into making prediction mistakes.
In 2014, Goodfellow \etal~\cite{goodfellow2014explaining} showed that an adversarial image of a panda could easily fool GoogLeNet \cite{szegedy2015going} into labeling it as a gibbon with high confidence, resulting in the birth of \emph{adversarial machine learning} (AML) research. 
Similar attack methods have been shown to achieve high misclassification rates in road sign classifiers \cite{eykholt2018robust} and evade automated surveillance cameras \cite{thys2019fooling}.
Though more and more people are studying and applying ML, many remain uninformed about the dangers of adversarial attacks to their models due to a lack of knowledge in AML.
As a result, the models developed often achieve good natural accuracy but are highly susceptible to attack-perturbed inputs \cite{sun2018survey}.
For these users (\eg, students, novice ML developers) to design or calibrate models to be adversarially robust for real-world applications, \re{it is essential to educate them about the concepts and impacts of adversarial attacks.}

\re{Many studies have shown that visualizations serve as effective educational tools for teaching complex ML concepts to non-experts interactively,} augmenting passive learning experience (\eg, textbooks and videos) \cite{wang2020cnn, kahng2019does, hohman2019s}.
Specifically, \re{we aim to design 
an educational visualization tool} to benefit learners who have an ML background but are unfamiliar with the risks of adversarial attacks, and are interested in learning AML to seek to build safer models for their applications.
For this work, we focus on evasion attacks in image classification, a highly active AML research path that most existing work \cite{goodfellow2014explaining, ilyas2019adversarial, zhang2019theoretically} focuses on since such models are frequently used in safety-critical applications \cite{guo2019survey, kuutti2020survey}. 
\re{Compared to adversarial attacks in certain other ML tasks such as NLP \cite{zhang2020adversarial} and recommender systems \cite{christakopoulou2019adversarial}, the perturbations applied to images also tend to be more human-imperceptible, making them even more challenging to understand and thus increasing the value of visualizing them for education. }

There are certain key challenges in understanding adversarial attacks for image classification. 
First, comprehending the attack process requires more than mere image inspection, as adversarial attacks utilize perturbations that exploit data features beyond human interpretation \cite{ilyas2019adversarial}.
These subtle modifications appear as imperceptible noise to human observers, making the adversarial images almost indistinguishable from their clean versions.
Second, an attack's efficacy varies based on the targeted instance and its label \cite{benz2021robustness}, meaning a few instances do not reflect the attack's behavior on the whole dataset or other classes, requiring multi-level inspection.
Third, the attack impact also depends on model architecture and training method \cite{madry2017towards, ilyas2019adversarial}, necessitating model comparison to understand attack variability as one model's performance cannot reflect the same attack's effects on others. 
Lastly, different attack methodologies yield varying impacts on models \cite{zhang2019theoretically, chen2017zoo}, so evaluations with different attacks and classifiers are needed to fully grasp the attack landscape. 
Therefore, visualizations need to be deliberately designed to illustrate the characteristics and effects of adversarial attacks, covering visual explanations of the attack logic, multi-level inspection across datasets and individual instances, model comparison to understand variability, and support for diverse attack methodologies.

However, existing \re{visualization-based educational tools} for AML fail to address these challenges, lacking comprehensiveness and generalizability in presenting various attack properties. 
For instance, Adversarial-Playground \cite{norton2017adversarial} is limited by its simplistic approach that displays an adversarial image beside its original, a method that is ineffective when the two images look identical from subtle perturbations.
Bluff \cite{das2020bluff} relies on visualizing the internal neuron logic on benign and adversarial examples, sacrificing model generalizability.
Both tools, limited to specific models/attacks, insufficiently represent evasion attacks by neglecting the influence of varying model architectures, training methods, and instance differences, leaving learners with an incomplete understanding.
While advanced AML visual analytic tools exist (\eg, AEVis \cite{cao2020analyzing} and Ma et al.'s \cite{ma2019explaining}), they are designed for experts to perform model analysis with complex visualizations that are challenging for novices to understand, thus not suitable for educational purposes. 
Further, AEVis lacks model comparisons and dataset-level visualizations, while Ma et al.'s is limited to data poisoning attacks in binary classification, lacking support for evasion attacks in multiclass classification.

\re{As such, designing an AML visualization targeted at novices for educational scenarios presents its unique challenges. 
These include effectively presenting advanced ML concepts like adversarial attacks in a non-overwhelming manner without losing essential details (\eg, the ``imperceptible'' attack property), along with creating an interactive learning environment that is comprehensive yet intuitive to understand. }
Therefore, to better augment learners' experience and address the limitations of existing tools, we carried out a study \re{to design an interactive educational visualization} to help learners understand evasion attacks at multiple levels, while allowing observation of their impacts on different models. 
Our primary objective is to help novice \emph{learners} gain a comprehensive understanding of the properties and risks of adversarial attacks from multiple lenses, thus enabling them to make more informed decisions during model development to mitigate the risks posed by adversarial attacks. 
Through this work, we have made the following contributions:

\begin{itemize}
    \item We conducted a \textbf{design study} on employing interactive visualization \re{for educational purposes} to augment the learning experience for AML. 
    Our study involved both literature reviews and user interviews with AML learners (N=3) and teachers (N=3) for design guideline formulation, followed by system development and an extensive evaluation.
    \item We designed and implemented \textbf{\name{}}, \re{an interactive educational visualization} for novice learners to gain a comprehensive understanding of adversarial attacks.
    To the best of our knowledge, \name{} is the first multi-faceted visualization designed specifically to support comprehensive learning of evasion attacks on both instance and population levels.
    Additionally, it supports model comparison and can readily adapt to different image classifiers and evasion attacks, addressing the generalizability gap in existing works (\eg, \cite{norton2017adversarial, das2020bluff}).
    \item We performed a \textbf{two-part evaluation} with 12 novice learners and 7 AML experts/teachers to quantitatively and qualitatively evaluate the learning aspects and usability of \name{}.
    Our results show that AdvEx not only is highly effective in facilitating understanding of adversarial attacks, but also offers an engaging and enjoyable learning experience, thus amplifying its educational impact.
    The strengths and limitations of \name{} are discussed, providing in-depth insights on how such a tool can be effectively utilized in an educational setting.
    
\end{itemize}

\section{Related Work} \label{related_work}

\subsection{Adversarial Machine Learning}

Many adversarial attacks have been proposed to work under different threat models, namely white-box and black-box attacks. 
A white-box attack has full access to the model's internals, while a black-box attack can only access model inputs and outputs.
Fast Gradient Sign Method (FGSM) \cite{goodfellow2014explaining}, Basic Iterative Method (BIM) \cite{kurakin2018adversarial}, and Projected Gradient Descent (PGD) \cite{madry2017towards} are some of the well-known white-box attacks. 
Advanced black-box attacks include Zeroth Order Optimization (ZOO) \cite{chen2017zoo}, HopSkipJump Attack \cite{chen2020hopskipjumpattack}, and Substitute Model Attack \cite{papernot2017practical}. 
To counter adversarial attacks, various defense methods have been proposed to fortify model robustness against adversarial inputs. 
The most effective defense is \emph{adversarial training}, which trains classifiers with adversarial examples by adding them to the training set \cite{goodfellow2014explaining, madry2017towards} or through regularizations \cite{zhang2019theoretically, qin2019adversarial}. 
TRadeoff-inspired Adversarial DEfense via Surrogate-loss minimization (TRADES) \cite{zhang2019theoretically} is a state-of-the-art adversarial training method that leverages a regularized surrogate loss from the observed trade-off between robustness and accuracy. 
Other examples of adversarial defenses include standard adversarial training \cite{madry2017towards}, robust self-training (RST) \cite{raghunathan2020understanding}, local linear regularization (LLR) \cite{qin2019adversarial}, etc.

\re{While \name{} can be employed with any evasion attack algorithm of the user's choice, to showcase the system’s adaptability to different types of attacks,} in this paper, we demonstrate \name{} using two attack examples recommended by the AML instructors we consulted, including FGSM, one of the earliest and most well-known white-box attacks \cite{goodfellow2014explaining}, and ZOO, a highly effective black-box query-based attack \cite{chen2017zoo}.
Several prior studies have tried to understand the characteristics of these two attacks. 
For example, Zhang et al. \cite{zhang2019defense} discovered that FGSM may create not only 2-D adversarial images but also 3-D adversarial examples by applying the attack methodology to PointNet \cite{Qi_2017_CVPR}, a DNN designed for 3-D point cloud data. 
Ye et al. \cite{9893902} applied adversarial attacks to a DL-based multiuser OFDM detector \cite{ye2017power} and showed that ZOO achieved the best performance among black-box methods.
Additionally, both attacks are frequently used in existing works to evaluate the effectiveness of adversarial defenses \cite{madry2017towards, he2019parametric, zantedeschi2017efficient, su2018robustness, 9282889} or as comparisons to other attacks \cite{8601309, moosavi2016deepfool, lin2022boosting}.
The abundance of existing works on both methods shows that they are well-known attacks and hence good introductory examples for those new to AML.
As AML is a relatively new area of ML, it is crucial to raise awareness on attacks like FGSM and ZOO to encourage users to build safer AI applications, especially those that are safety-critical. %

\subsection{Visualizations of Adversarial Attacks}

In general, interactive tools designed for visualizing adversarial attacks are relatively under-explored. 
A few tools with educational purposes have been proposed in past studies.
Adversarial-Playground \cite{norton2017adversarial} is a simple web application that demonstrates the efficacy of three attack algorithms against a small CNN on the MNIST dataset \cite{deng2012mnist}. 
The tool allows users to choose from a set of pre-defined inputs and displays the adversarial image next to its original alongside classification likelihoods to illustrate the attack. 
Bluff \cite{das2020bluff} visualizes attacks on a vision-based network, but focuses on model internals instead by highlighting the neurons and connections that an attack exploits to confuse the model.

However, these tools lack comprehensiveness and multi-faceted approaches in visualizing adversarial attacks.
For instance, Adversarial-Playground \cite{norton2017adversarial} offers a simple image comparison approach that becomes ineffective if used to visualize attacks that generate ``imperceptible'' inputs, a common characteristic among adversarial attacks.
Its applied perturbations on the black and white MNIST dataset are also highly visible, which could create a false sense of security among learners about their abilities to discern adversarial images from clean ones.
Bluff \cite{das2020bluff}, on the other hand, relies on visualizing a model's internal neurons, thus not easily extendable to other model architectures.
Both tools are restrained to specific attacks/models, supporting visualization of only one classifier and several instances, missing dataset-level attack information and inadequately demonstrating the variability in attack impacts due to differences in model structures and training methods.

In addition, a few advanced AML visual analytics tools have been developed as well. 
AEVis \cite{cao2020analyzing} uses a river-based visual metaphor to show how the datapaths of clean and adversarial examples merge or diverge within the network.
However, it suffers from the same limitation of lacking model comparisons and dataset-level information, and cannot be used to visualize attacks with varying perturbation sizes.
Ma et al. \cite{ma2019explaining} proposed a framework that employs a multi-level visualization scheme to support the analysis of data poisoning attacks in binary classification tasks.
While comprehensive, it is designed specifically for data poisoning attacks in binary classifications, thus diverging in focus from this work and lacking support for evasion attacks in multiclass classifications.
Moreover, both tools are designed primarily for experienced practitioners to perform visual analytics on models under adversarial attacks, featuring complex visualizations and interfaces that may be overwhelming for novice learners.

Hence, current educational tools lack comprehensiveness, often visualizing a few instances and limited to specific attacks and models; current advanced visual analytics tools are overly complex for our intended audience or have a different focus from this work.
In contrast, with \name{}, we aim to enable users who have little or no knowledge of AML to learn about adversarial attacks at both dataset and instance levels, while making it easy to be generalized to different evasion attacks and vision-based classifiers.

In addition, to visualize the shift in how models perceive the dataset before and after an attack, we incorporated a dimensionality reduction overview depicting each model's feature space in \name{}.
Dimensionality reduction has been used frequently to understand and visualize adversarial attacks. %
For instance, Ma et al. \cite{ma2019explaining} and Park et al. \cite{park2021vatun} utilize t-SNE for data embedding views to visualize the impacts of data augmentations including adversarial attacks.
Panda and Roy \cite{panda2021implicit} introduced a Noise-based Learning (NoL) approach for training robust DNNs and provided simplistic PCA-based visualizations for adversarial dimensionality and loss surface visual analysis.
Hendrycks and Gimpel \cite{hendrycks2016early} incorporated PCA into adversarial image detection and visualized how adversarial images abnormally emphasize coefficients for low-ranked principal components.
Inspired by these works, in \name, we apply similar methods to project the data embeddings onto a 2-D plane, and use animated transitions and colors of circular glyphs to visualize how the attacks alter the models' perception of the images.

\subsection{Visualizations for Learning ML}

Outside of AML, several visualization tools specifically designed for learning ML have been proposed as well. 
GAN Lab \cite{kahng2019does} is designed for non-experts to learn and experiment with generative adversarial networks (GANs) by visualizing GANs' dynamic training processes on a simple dataset. 
CNN Explainer \cite{wang2020cnn} enables learners to inspect the interplay between CNNs' low-level mathematical operations and their high-level model structures. 
Summit \cite{hohman2019s} provides higher-level explanations of DNNs by visualizing image features detected by the networks and how those features interact to make predictions. 
\ree{More recently, TransforLearn \cite{gao2023transforlearn} provides interactive visual tutorials for learners to understand transformer models by supporting architecture-driven and task-driven exploration.}
While these tools are effective for demonstrating basic ML concepts, they are not suitable for our study's design objective in the context of AML learning.
For example, GAN Lab \cite{kahng2019does} is only for exploring generative models on low-dimensional training datasets and significantly diverges from the focus of this work. 
\ree{Similarly, TransforLearn \cite{gao2023transforlearn} is specifically designed for learning transformer models' layer operation and mathematical details.}
While CNN Explainer \cite{wang2020cnn} and Summit \cite{hohman2019s} could potentially be extended to explore a model's internal datapaths on adversarial examples, they would still share the limitations of lacking model generalizability and dataset-level attack information like Bluff \cite{das2020bluff} and AEVis \cite{cao2020analyzing}. 
\re{As such, it is important to have a designated educational tool for AML that addresses the gaps of existing works. 
We selected AML as our target domain as it is crucial to educate novice practitioners who apply ML across diverse domains but do not understand their models' vulnerability due to their gap in AML knowledge. 
We believe by helping them understand adversarial attacks in a hands-on manner, they would be equipped with the necessary knowledge to design models that are robust for real-world applications in the future. }

Despite focusing on visualizing common DNNs instead of adversarial attacks, all aforementioned studies have provided us with inspirations for \name{}'s design. 
Specifically, similar to GAN Lab \cite{kahng2019does} and CNN Explainer \cite{wang2020cnn}, \name\ is accessible to any user with a modern browser without the need to install specialized hardware for deep learning. 
Motivated by GAN Lab \cite{kahng2019does}'s step-by-step training visualization, \name\ provides step-by-step executions of the attack methodology to visualize the detailed attack process. 
Like CNN Explainer \cite{wang2020cnn} and Summit \cite{hohman2019s}, \name\ also adopts smooth transitions across different levels of abstraction to facilitate visual exploration and to serve as the link that connects different views of the visualization tool. 
\ree{Inspired from TransforLearn \cite{gao2023transforlearn}, \name\ uses task-driven exploration to help users gain a deeper understanding of model robustness with actual image classification tasks. }
Based on existing work, we aim to develop \name{} as a tool with comprehensive visualizations and animations that can enable intuitive exploration of attack properties across multiple levels.
\section{Design Goals}
To formulate the design guidelines for \name{}, we conducted user interviews with six participants, including three interviewees (S1, S2, S3) who have AML learning experience and three AML teachers (E1, E2, E3). 
Our goal was to understand learners' needs in understanding adversarial attacks and to have experienced AML teachers envision how such a tool can be utilized in an educational setting.
The learners involved come from computer science and data science backgrounds, and their employed learning methods varied from enrolling in AML courses to reading academic papers or online blog posts. 
The teaching experience of the interviewed educators ranged from leading graduate-level AML seminars to overseeing AML components within undergraduate ML courses.
The semi-structured interviews lasted between 60 to 90 minutes and covered the following topics: 1) the participants' background and experience in AML learning/teaching, 2) existing content or tools used to understand/teach AML, 3) the challenges in understanding/teaching adversarial attacks, 4) features and functionalities to include in an educational visual tool for adversarial attacks, and 5) how participants envision using such a tool in an educational setting. 
The participants were compensated $\$$20/hour for the interview.

While none of the interviewees had previously used any visualization tool for adversarial attacks, all recognized the value of introducing a multi-level visualization tool to demonstrate evasion attacks to learners.
Specifically, they believed that an interactive visualization tool would have multiple educational benefits, including \pqt{providing an accessible way to demonstrate attacks in practical applications}{E2}, \pqt{making the learning experience more engaging}{E3}, and \pqt{accommodating learners with different backgrounds}{E1}. 
The interviewees also thought that the tool could be used either in a self-learning scenario for exploration or incorporated into AML courses to demonstrate concepts and better augment students' learning experience. 
These comments confirm the need for a visualization tool like \name{} in both independent and guided AML learning contexts.

We transcribed our interviews and employed a hybrid method of open and closed coding, informed by an extensive literature review (\autoref{related_work}), to analyze the gathered qualitative data. 
Using an affinity diagram, we identified recurring themes and requirements of such a visualization tool for AML learning. 
\re{This process involved collaboratively organizing observations and insights from literature and transcriptions into sticky notes, which we grouped on a large canvas based on their similarities and common themes. 
We started off by using open coding to freely identify themes and patterns in our data. 
Then, as the themes became clearer, we organized them into a more structured framework. 
Through iterative sessions of discussion and reorganization, clusters of notes representing common themes/requirements for our AML visualization were formed.}
As a result, we derived the following design goals to guide the development of \name{}:

\begin{itemize}
    \setlength\itemsep{0mm}
    
    \item[\textbf{G1}] \textbf{Present visual abstraction of the attack impact at multiple levels.} 
    Many existing tools (\eg,\cite{norton2017adversarial,das2020bluff}) only display instance-level attack information, such as how a specific image is modified by the attack. 
    These instance details are insufficient to illustrate the reason behind misclassifications or the overall attack impact on a larger dataset. 
    E3 mentioned, \qt{When simply comparing the images, we can observe the differences from a human perspective, but it remains unclear why the model misclassifies them.}
    E2 agreed that \qt{Examining images prone to misclassification is vital, but seeing the broader impact is equally important to fully grasp the risks.}
    Therefore, visual abstractions at multiple levels should be included to provide both dataset-level overviews of the attack and the options to conduct more in-depth investigations on specific instances.

    \item[\textbf{G2}] \textbf{Design a visualization framework that can be generalized to different evasion attacks and image classifiers.}
    Generalizability is crucial as it enables learners to grasp the variability of attack methods, assess different kinds of models under attacks, and connect theoretical knowledge with practical applications.
    E3 confirmed that \qt{A key learning objective should be the various methods to generate adversarial examples, which is essential for understanding how to defend against these diverse attack strategies.}
    \re{S2 and S3 agreed that generalization can help learners gain practical insights into the variability of attack impact by exploring different attacks in actions and visualizing models/attacks that align with their backgrounds, rather than being restricted to a predetermined set of models/attacks.}
    For \name{}, we aim to address the gap of existing works \cite{norton2017adversarial, das2020bluff} being constrained to specific attacks/models by designing a general framework in these aspects to enable a more holistic and practical understanding of the attacks. 
    \re{Specifically, we aim to adapt a ``plug-and-play'' approach to allow users to easily swap out the attack algorithms and models based on their interests and learning goals. }

    \item[\textbf{G3}] \textbf{Enable comparative analysis of different models' robustness under attack.} 
    Models with different architectures and training methods vary in their robustness against the same attack \cite{zhang2019theoretically, ilyas2019adversarial, goodfellow2014explaining}, but most learning tools \cite{norton2017adversarial, das2020bluff} demonstrate attacks with a single, arbitrary model.
    Enabling visual analysis of multiple models is important to facilitate understandings of the variability in attack impact and to highlight the rationales for why certain models would fail. 
    E3 stated, \qt{Comparing the model differences provides insights into why certain attacks succeed or fail, going beyond just seeing changes in model accuracy.}
    E2, S2, \& S3 agreed that model comparisons \pqt{highlight the models' varying defense abilities}{S3} and in turn \pqt{helps learners defend and improve their own \re{in their future applications}.}{E2.}
    Thus, we aim to provide side-by-side comparisons of different models under various attack scenarios.

    \item[\textbf{G4}] \textbf{Facilitate dynamic experimentation with fluid transition between different perturbation sizes.} 
    As the perturbation size increases, the attack becomes more effective and the applied noise also becomes more visible.
    Allowing users to dynamically experiment with the perturbation size and observe the changes in real time \pqt{facilitates a better understanding of this correlation}{E2} and \pqt{creates a more game-like, engaging learning process}{S1}.
    E1 pointed out that such experimentation \qt{allows learners to observe how the model's perception of an instance changes, and identify the threshold at which misclassification occurs.}
    S3 stated that the approach \qt{helps learners understand when exactly the image starts to look different for humans.}
    Therefore, interfaces are included to allow easy manipulation of the perturbation size and visualize the changes in real time. 

    \item[\textbf{G5}] \textbf{Allow step-by-step execution for learning the attack process in detail.} 
    Mentioned by E1, E2, \& S2, navigating complex mathematical steps in papers to understand attack logic is a daunting task for learners.
    A step-by-step attack execution \pqt{provides a more structured understanding of attack strategies}{E2}. 
    This approach allows learners to \pqt{grasp not just the impact but also the design and rationale behind the attacks}{E3}.
    E2 confirmed, \qt{A step-by-step approach simplifies the attack process and reduces learners' burden compared to interpreting steps directly from papers.}
    In \name{}, we aim to incorporate a step-by-step view to clarify the underlying attack logic for learners, guiding them through the complexities of various attack strategies.

\end{itemize}

\newcommand{\xv}{\mathbf{x}}
\newcommand{\nv}{\mathbf{n}}

\section{\name}

\begin{figure}[tb]
\centering
\includegraphics[width=0.8\columnwidth]{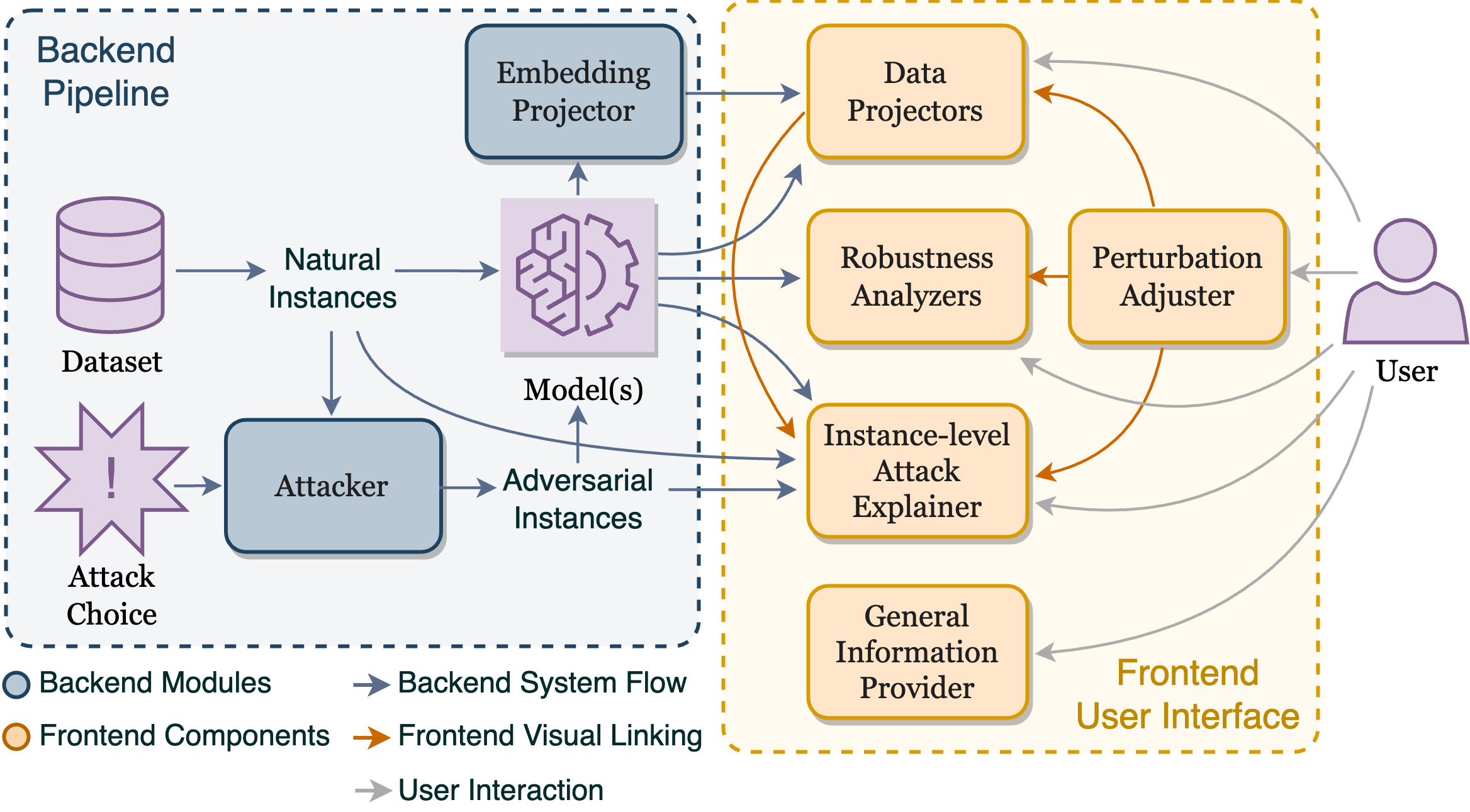}
\caption{A schematic diagram depicting the system architecture of \name{}. 
In the backend pipeline, an Attacker module performs users' choice of attacks on the image dataset, targeting models specified by users (\textbf{G2}). Once processed, the backend outputs are passed to the frontend interface for user interaction. }
\label{fig:schematic_diagram}
\end{figure}

Based on our design guidelines, we developed \name{}. 
Here, we begin with an overview of \name{}'s system, followed by detailed descriptions of its backend modules and frontend components.

\subsection{System Overview}

As depicted in \autoref{fig:schematic_diagram}, \name{} is a web application with two main system components: A) a \textit{backend pipeline} (\autoref{backend}) and B) a \textit{frontend user interface} (\autoref{frontend}).

In the backend pipeline, an \textit{Attacker} module begins by normalizing the image data and performing users' chosen attack methods to generate adversarial examples. 
Both the original and adversarial examples are fed into the models to obtain information such as image embeddings, confidence scores, and prediction accuracy.
An \textit{Embedding Projector} is employed to extract each model's embedding vectors by removing the final output layer and applying dimensionality reduction methods (\eg, t-SNE \cite{van2008visualizing}, PCA \cite{pearson1901liii}) to prepare the projection coordinates of the data representations.
The processed outputs are relayed to the frontend components to be presented visually for user interaction.

The frontend interface comprises five key components: 1)  \textit{Data Projectors} (\autoref{fig:teaser}c), 2) \textit{Instance-level Attack Explainer} (\autoref{fig:teaser}d), 3) \textit{Robustness Analyzers} (\autoref{fig:teaser}a), 4) \textit{Perturbation Adjuster} (\autoref{fig:teaser}b), and 5) \textit{General Information Provider} (\autoref{fig:teaser}e) + interactive tutorials. 
The Robustness Analyzers feature two interactive bar charts that assess the models' overall robustness under a specified attack (\textbf{G1}) and offer a comparative view of this robustness to natural accuracy (\textbf{G3}). 
The Data Projectors utilize coordinates from the Embedding Projector to visualize data representations as two interactive, side-by-side scatterplots.
These scatterplots enable exploration of attack-induced embedding changes (\textbf{G1}) and offer comparisons of embeddings between different models (\textbf{G2, G3}).
The Instance-Level Attack Explainer provides detailed insights into a specific instance (\textbf{G1}), complemented by a confidence score view and a step-by-step guide to the instance's attack process (\textbf{G5}). %
The Perturbation Adjuster allows users to select their desired perturbation size and initiates animations within the three aforementioned components to simulate the attack in real time (\textbf{G4}).
Finally, along with interactive tutorials, the General Information Provider guides users through the navigation of the interface and offers further context on AML.

\subsection{Dataset and Models}

In this paper, we use the CIFAR-10 dataset \cite{krizhevsky2009learning} to demonstrate \name{}, but our system can be employed with any image dataset with $\le$ 12 classes \re{due to color distinguishability} \cite{munzner2014visualization} or a subset of a dataset with more classes. 
The CIFAR-10 dataset consists of 60,000 \(32 \times 32\) colored images from 10 different classes (50,000 training data and 10,000 testing data), with 6,000 images per class. 
We chose this dataset due its popularity of being used in ML research to evaluate the accuracy and robustness of image classifiers  \cite{zhang2019theoretically, croce2020reliable, hendrycks2019using}.

In addition, \name{} supports a variety of image classifiers and allows the user to compare two models side by side (\textbf{G2, G3}).
For example, users could compare CNNs with the same architecture but different numbers of convolutional layers, or investigate how a classifier trained adversarially may outperform a standard model in an attack. 
For this paper, we loaded two pairs of models for our studies: 1) VGG-16 vs. VGG-19, and 2) ResNet-34 trained naturally vs. trained adversarially with TRADES \cite{zhang2019theoretically}. 

\subsection{Backend Pipeline} \label{backend}
In this section, we describe how the backend processes and analyzes the data in \name{}, including how it generates the adversarial examples and prepares the data instances and model outputs for frontend display.

\subsubsection{Attacker Module}
The ``Attacker'' module produces adversarial examples of the original dataset by conducting adversarial attacks on the targeted models. 
It first feeds the natural images into the targeted models (or surrogate models) to obtain information relevant to the attack, then adjusts the pixel values of the input image based on the information.
\re{While users can swap out the attack algorithm in the Attacker module with any evasion attack they wish to learn about (\textbf{G2}), here we use one white-box attack and one black-box attack, FGSM \cite{goodfellow2014explaining} and ZOO \cite{chen2017zoo}, as examples for demonstrating our system's adaptability to different attacks. }

We chose the FGSM attack due to its notoriety for creating the very first adversarial panda image \cite{goodfellow2014explaining} that is well-known among AML researchers.
It is commonly used as a baseline for evaluating model robustness and defense effectiveness  \cite{madry2017towards, zhang2019defense, rathore2020untargeted}.
The attack was also recommended by the consulted AML instructors as it is relatively simple in logic and used as the introductory attack in AML courses and tutorials. 
However, the attack has been proven to be extremely effective \cite{goodfellow2014explaining}:

\begin{equation}
\xv' = \xv +\epsilon \textrm{\textrm{sign}}(\nabla_{\xv} J(\theta, \xv, y)). 
\end{equation}

It modifies image $\xv$ by maximizing the loss $J(\theta, \xv, y)$ towards the gradients' sign to produce the adversarial image $\xv'$. 
Here, $y$ is the true label, $\theta$ is model parameters, and $\epsilon$ scales the perturbation. 
We used $L^{\infty}$ norm to restrict the maximum pixel change to create bounded examples. 

Additionally, we employed the ZOO attack \cite{chen2017zoo}, an advanced black-box attack, as the second demonstration method. 
The AML instructors highlighted that ZOO is often used as a representative example of black-box attacks in AML courses.
It is also frequently used to evaluate defenses \cite{9282889, he2019parametric} or as a benchmark for other attacks \cite{lin2022boosting, ye2017power}.
ZOO only has access to model inputs (e.g., images) and outputs (e.g., confidence scores), and finds $\xv'$ by solving the following optimization problem:

\begin{equation}
\begin{aligned}
& \underset{\xv'}{\text{minimize}}
& & \lVert \xv' - \xv \rVert_2^2 + c \cdot f(\xv') \\
& \text{subject to}
& & \xv' \in [0, 1]^p
\end{aligned}
\end{equation}

The first term $\lVert \xv' - \xv \rVert_2^2$ applies $L^{2}$ norm regularization to enforce similarity between $\xv'$ and $\xv$. %
The loss $c \cdot f(\xv')$ represents the level of unsuccessful attacks, with $c > 0$ as the regularization parameter.
The attack approximates the gradient with a finite difference method and solves the optimization problem via zeroth order optimization.

Employed with users' selected attack, the module performs attacks respectively with the selected perturbation sizes $\epsilon$.
Based on \textit{RobustBench}'s suggested limits \cite{croce2020robustbench}, we selected $\epsilon$ of 0.00, 0.01, 0.02, and 0.03 for FGSM with $L^{\infty}$ norm, and $\epsilon$ of 0.0, 0.1, 0.3, and 0.5 for ZOO with $L^{2}$ norm for our demonstration.
The resulting adversarial examples, along with the original data, are then inputted in the models for classification and embedding extraction.

\subsubsection{Embedding Projector}
The Embedding Projector is tasked with 1) processing the models' produced embeddings and 2) analyzing the information of the extracted features and preserving it in a low-dimensional representation.
The goal is to unveil important patterns in the embeddings and transform them into a format easily fetched by frontend.
The module temporarily detaches the final output layer to obtain the embeddings and reduces their dimensions by applying users' choice of dimensionality reduction for later 2-D visualizations.
For instance, in the case of t-SNE, the module analyzes instance features by constructing a lower-dimensional probability distribution that represents the similarities between the objects in the high-dimensional space. 
If PCA is used, the module preserves the most significant variability in the embeddings while reducing the number of features.
The resulting outputs are scaled to be used as the x- and y-coordinates of the instances in scatterplots and are stored as tabular data easily accessed by the frontend Data Projectors.

\subsection{Frontend User Interface} \label{frontend}

Here, we detail the frontend components of \name{}.
We demonstrate our approach using FGSM on VGG-16 and VGG-19 models pre-trained with CIFAR-10 \cite{huyphan_2021}.

\subsubsection{Data Projectors} 

\begin{figure}[tb] 
\centering
\includegraphics[width=0.8\columnwidth]{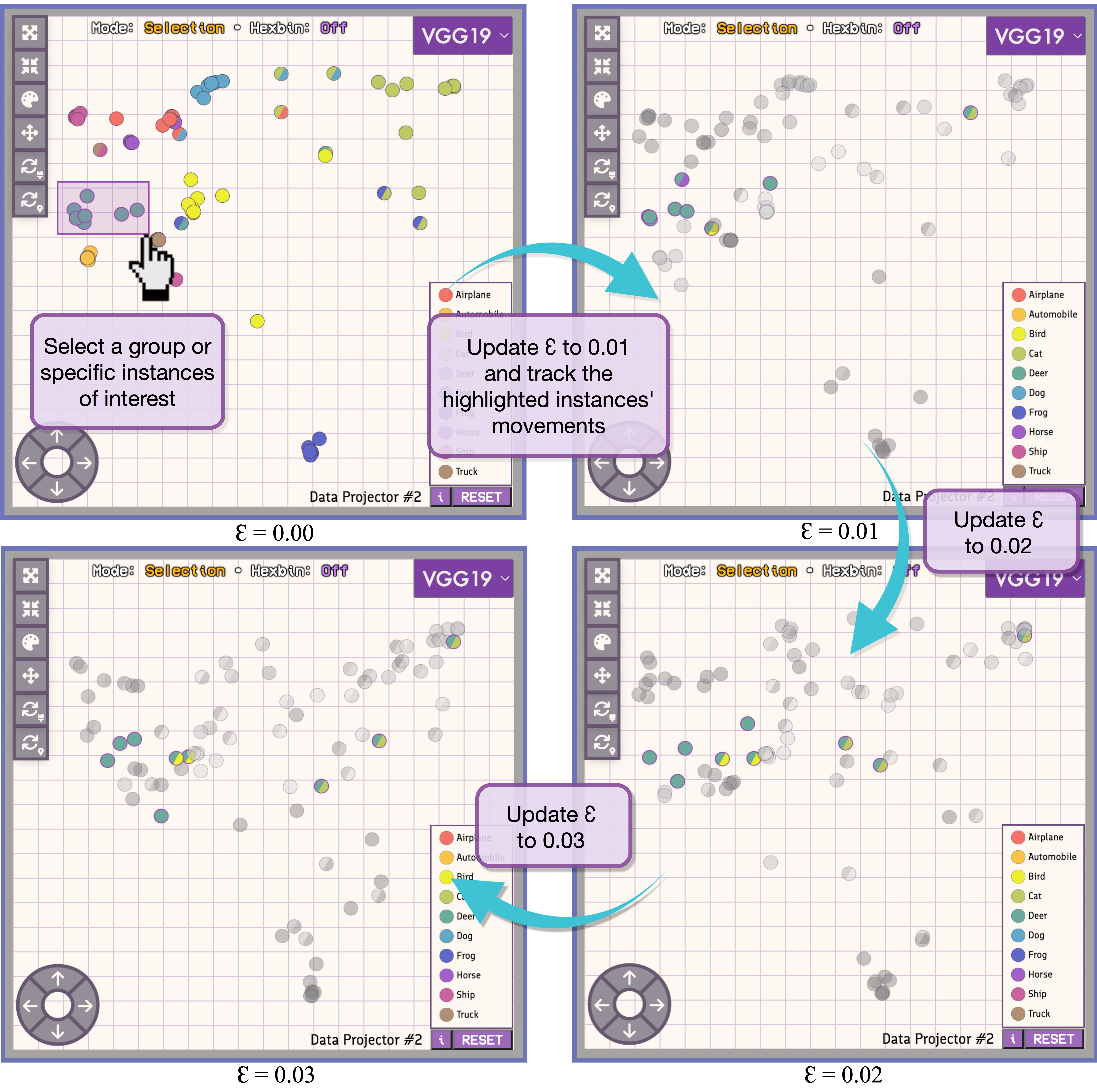} 
\caption{A user highlights and tracks a specific class from the dataset with selection mode. Under this mode, one can evaluate model performance on a dataset subset.} 
  \vspace{-5mm}
\label{fig:brush}
\end{figure}

\begin{figure}[tb] 
\centering
\includegraphics[width=\columnwidth]{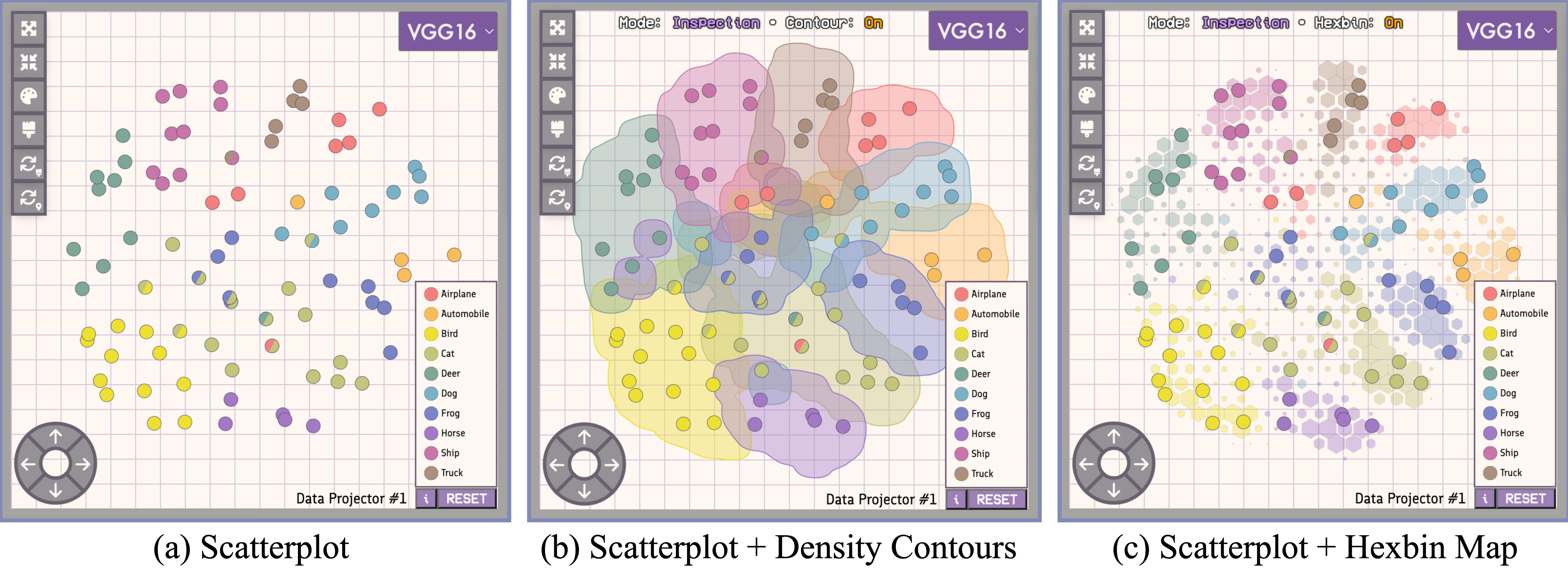} 
\caption{We explored a variety of visual encodings and aggregating features for the Data Projectors. We chose binned aggregation with multiple zoom levels, with an optional hexbin toggle to display the overall distribution (Fig. c). This preserves data scalability and displays global data structure without the need for high-performance devices.} 
  \vspace{-5mm}
\label{fig:hexbin}
\end{figure}

The Data Projectors (\autoref{fig:teaser}c) represent dimensionality reduction overviews of the dataset and consist of two scatterplots where the image embeddings are projected as circles on a 2-D plane. 
Each circle corresponds to a data instance and is sliced into two halves: the color of the left half represents the instance's ground truth label, while the color of the right half represents its current prediction. 
The spatial positions of the circles encode the relationships between them in the original high-dimensional space (\eg, similarities, variance, local and global structure). 

\re{
Inspired by \textit{nanocubes} \cite{lins2013nanocubes}, we use a combination of binned aggregation and hierarchical clustering with multiple zoom levels to preserve data scalability and organize embeddings into a multi-level structure (\autoref{fig:teaser}c1). 
Each level of data cubes represents varying granularity of data aggregation, with the reduced space split up into a grid of equally shaped and sized bins. 
The number of bins at the highest level is scaled based on the data range in reduced dimensions, and follows a fixed multiplying pattern for subsequent levels (\eg, 10 $\times$ 10, 20 $\times$ 20, 30 $\times$ 30, ...). 
Data points are allocated to bins based on their positions in the scatterplot, and a representative instance, predicted as the most frequently predicted class within that bin, is sampled for display at that specific level. }

\re{To assist users in identifying data patterns at higher zoom levels with fewer sampled instances, the size of the hexagons in the background indicates the density of embeddings in each bin (\autoref{fig:hexbin}c), visualizing the overall clustering patterns.  
During exploration, the projectors dynamically sample and display instances from each level, allowing for an iterative and detailed examination of larger datasets without getting messy from too many moving dots.}
Our approach allows interactive exploration of data sources with large numbers of instances while maintaining the global data structures without high-performance devices. 

When an attack is conducted with newly specified magnitude, the Data Projectors visualize the attack with an animated sequence that emphasizes each circle's change in position and color  (\textbf{G4}) \re{(\autoref{fig:brush})}. 
For example, if a circle transitions to a different coordinate, this indicates that the model's perception of the instance's features has been altered by the attack. 
\re{To mitigate potential artifacts produced by projection methods' randomness, multiple runs of the projection are conducted and the results are averaged. 
This approach allows us to ensure that the transitions observed in the projectors are predominantly indicative of the models' changes in feature perceptions. }
Moreover, if the class ``airplane'' is represented by the color red and the class ``automobile'' is represented by the color orange, then a red circle transitioning into a half-red, half-orange circle means that this is an airplane image incorrectly classified as an automobile due to the attack. 
To improve the usability of the projectors, the following functionalities are also incorporated:

\begin{itemize}
    \item \textbf{Inspection mode. \wsicon{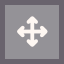}}
    Under this mode, users can zoom and drag freely within the scatterplots to explore the models' embedding distributions.
    To avoid overlap when instances share similar features, \name{} dynamically adjusts the radius of the projected circles at different zoom levels, allowing users to precisely examine each individual instance.
    Clicking a circle highlights the instance by enlarging its radius and pinning it, then panning the entire plot to recenter that circle within the 2-D plane.

    \item \textbf{Selection mode (\autoref{fig:brush}). \wsicon{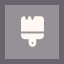}} 
    In this mode, users can highlight a subset of the dataset, including a single item, by specifying a selected region with a pointing gesture. %
    As a result, only colors of the selected circles within the region remain visible, while all other instances are grayed out. 
    This feature allows users to track the movements of specific subgroups or instances across different perturbation sizes, adding a subpopulation-level display (\textbf{G1}). 
    When a group/instance is highlighted in one Data Projector, the same group/instance is simultaneously highlighted in the other projector for comparison (\textbf{G3}). 

    \item \textbf{Hexagonal binning toggle. \wsicon{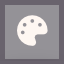}}
    While navigating, users can toggle a hexagonal binning map (\autoref{fig:hexbin}c) for each projector to track the global data structure.
    The hexbin map displays the general trends in instance clustering based on model predictions, allowing quick identification of decision boundaries and similarly classified image groups (\textbf{G1}).
    Moreover, this approach preserves visibility of the whole dataset's distribution even when the projectors are only displaying a subset at higher zoom levels.

\end{itemize}

In summary, the Data Projectors provide interactive visualizations of image embeddings, illustrating instance relationships via spatiality and revealing population-level attack impacts though animated transitions (\textbf{G1, G4}).
Given that the set of generated adversarial examples varies depending on the chosen attack method and the model targeted, we provide side-by-side visualizations of two distinct models' embeddings to illustrate the differential impacts of the attack (\textbf{G3}).
We design the Data Projectors to not only be technically robust to handle varying data scales and complexities, but also intuitively understandable through its animated transitions to visualize changes in model perception. 
Compared to works like \cite{norton2017adversarial}, our approach illustrates attacks' impacts on models while preserving their imperceptibility at the instance level (see \autoref{sec:instance_level_explainer}).
By interacting with the Data Projectors, users can intuitively observe how changes in the perturbation size influence both the models' data representations and their resulting image predictions, thus gaining a deeper understanding of the attack's impact on model performance.

\subsubsection{Instance-level Attack Explainer} 
\label{sec:instance_level_explainer}

\begin{figure}[tb] 
\centering
\includegraphics[width=0.8\columnwidth]{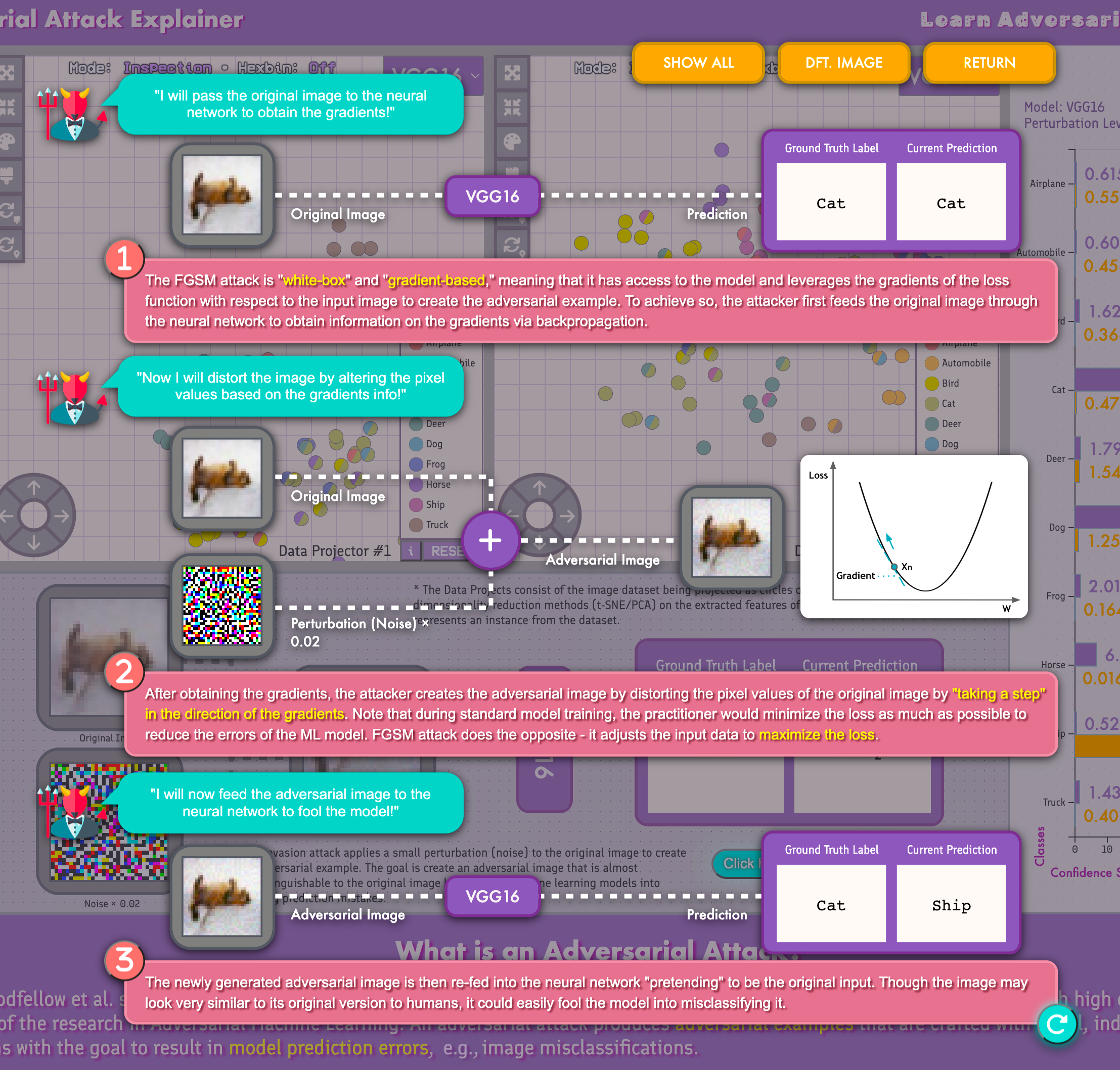} 
\caption{An example of the \re{final state} of the step-by-step execution view for explaining the FGSM attack. 
The view progressively reveals attack elements and explanations, animated one by one to illustrate the flow of the attack process.}
  \vspace{-5mm}
\label{fig:step}
\end{figure}

While the Data Projectors visualize attacks on dataset or subpopulation levels, the Instance-level Attack Explainer \re{(\autoref{fig:teaser}d)} provides in-depth information for each perturbed input. 
It outlines the attack process for each image, detailing instance-level information such as the original image, applied noise, and confidence scores (\textbf{G1}). 
To examine an instance, users click on the corresponding circle in the Data Projectors to update the panels associated with the attack explainer. 
Specifically, the Instance-level Attack Explainer consists of the following components:

\begin{itemize}

    \item \textbf{General view.} 
    The general view (\autoref{fig:teaser}d1) displays key information of the selected instance, including its original image, applied noise, adversarial image, targeted model, true label, and current prediction (\textbf{G1}). 
    To help users contextualize these instance-level details, subtle animations are used to link the information together to depict the high-level attack flow.
    For instance, a repeated animated sequence shows the original image and the generated perturbation progressively moving towards each other with reduced transparency and stacking on top of each other, then gradually fading into the final perturbed image. 
    Dashed lines connecting the images are animated to continuously move from the clean image + noise to the resulting adversarial image. 
    By presenting a visual attack narrative, these animations are designed to help users better interpret the instance-specific information within the context of the current attack.
    
    \item \textbf{Side-by-side image inspection.} 
    For a closer inspection of the images, users can click on the image thumbnails in the general view to view enlarged versions. 
    A comparison mode is available to examine the clean and adversarial images side by side and observe the exact pixel differences.
    
    \item \textbf{Confidence score view.} 
    The interactive bar chart panel (\autoref{fig:teaser}d2) showcases the model's pre- and post-attack confidence scores across all classes for the selected instance.
    These scores are grouped together in pairs to provide comparison between the model's confidence before and after an attack.
    Hovering over each pair of them reveals their exact difference in percentage, allowing users to quantitatively assess the attack impact on class-wise classification probabilities (\textbf{G1}).
    
    \item \textbf{Step-by-step execution view.} 
    The step-by-step execution view (\autoref{fig:step}) provides detailed explanations of the underlying attack logic. 
    Clicking the button at the bottom-right of the general view activates this feature, which initiates a series of step-by-step animated sequences with accompanying explanations (\textbf{G5}). 
    Once activated, these explanations unfold sequentially.
    For instance, Explanation \#2 (\autoref{fig:step}-2) does not appear until users click the play button next to Explanation \#1 (\autoref{fig:step}-1), which becomes visible only after Explanation \#1 has finished playing. 
    \re{Within each explanation, individual elements are animated to present the flow of the attack. 
    This includes animating the appearance of components one by one to show how the input image is fed into the model to obtain relevant information, or how the generated noise is added to the image to create the adversarial input. }
    A toggle allows users to replace the default image with their selected instance for the view's demonstration, allowing them to apply step-by-step explanations to an actual adversarial example they are examining.
    This feature provides users with a more tangible and personalized understanding of how adversarial attacks manifest and operate on real-world examples (\textbf{G2}).

\end{itemize}

In short, the Instance-level Attack Explainer offers a focused, in-depth look at adversarial attacks on individual instances  (\textbf{G1}).
\re{To achieve a balance between technical detail and novice engagement,} the view translates complex attack processes into intuitive visual narratives, and adopts a step-by-step approach to guide novices through the underlying attack logic (\textbf{G5}). 
Also, its confidence score view enables users to quantitatively explore and assess how the attack impacts the model's confidence for the given instance (\textbf{G1}).
Together, these features offer a detailed perspective of the instance-specific properties and consequences of adversarial attacks.

\subsubsection{Robustness Analyzers} 

The Robustness Analyzers \re{(\autoref{fig:teaser}a)} in the leftmost panel feature two compact, interactive bar charts, each containing two bars. 
These charts evaluate the model's robustness under the given attack and compare it to the pre-attack accuracy (\textbf{G1, G3}).
The left bars represent natural accuracy, indicating the model's prediction accuracy on the clean dataset, while the right bars represent robust accuracy, reflecting the model's performance on the adversarial dataset.
As users adjust the perturbation size \re{(\autoref{fig:teaser}b)}, the right bars dynamically adjust their heights to visualize the corresponding changes in the model's robust accuracy (\textbf{G4}).
With the Robustness Analyzers, users can compare 1) a model's robustness to its baseline performance and 2) the relative performance of different models under standard and adversarial conditions (\textbf{G2, G3}).
Consequently, users can gain insights into the attack's varying impact across models, identify which models are resistant or vulnerable to the current attack, and quantify the degree of performance degradation from adversarial inputs.

\subsubsection{Perturbation Adjuster} 
The Perturbation Adjuster \re{(\autoref{fig:teaser}b)}, situated below the Robustness Analyzers, features a slider and an attack button.
The slider allows users to choose a perturbation size from a range they have pre-set in the backend, which they can adjust horizontally to visualize the desired attack strength. 
\re{We chose the perturbation size as \name{}'s primary control parameter as in the context of AML, it is standard to evaluate attacks under a defined upper bound on the perturbation size. 
While different attacks vary in their logic, all have a perturbation size that can be defined when applied to input images, which is what we focus in \name{}.}
Upon selecting a perturbation size, users initiate an animated attack sequence by clicking the attack button, which triggers changes in other components of the interface.
For example, the circles of the Data Projectors' \re{(\autoref{fig:teaser}c)}  may shift to new coordinates alongside prediction color changes, while the right bars of the Robustness Analyzers \re{(\autoref{fig:teaser}b)} adjust their heights up or downward based on the model's accuracy with the new adversarial dataset.
With the Perturbation Adjuster, users can dynamically modify the perturbation size and observe the increased attack strength with larger perturbation sizes, as well as the growing visibility of applied image noise (\textbf{G4}).
The integration of dynamic perturbation control and real-time visual feedback enables users to intuitively understand the interplay between perturbation size, attack intensity, and resulting image distortions \re{across a variety of attack methods}.

\subsubsection{Interactive Tutorials + General Information Provider}

To help users pick up \name{} more easily, an interactive tutorial system is also integrated. 
Upon launching the application, users encounter an overlay tutorial that introduces every component of \name{}'s interface, highlighting its key features.
During interaction, hovering over any Data Projectors' button displays a tooltip explaining its function. 
If users have not engaged with certain key features (\eg, hexbin map, step-by-step execution) within 10 minutes, an animated arrow prompts them to explore these features.

Furthermore, if users wish to learn more about \name{} and AML research, they may read the information placed beneath the interactive components \re{(\autoref{fig:teaser}e)}, which provides more in-depth explanations for both.
By including interactive tutorials and reading materials, users will not only learn our tool faster, but also gain detailed and accurate knowledge of adversarial attacks in addition to perceiving them through interactive visualizations. 
\re{As such, we designed \name{} to place balanced emphasis on both visualizations and text, reinforcing novices' learning by presenting content in different formats, allowing learners to quickly grasp complex topics through multiple forms of interpretations. }

\section{User Study with Novice Learners}

To assess how \name{} can help novice AML learners, we conducted a user study with participants who knew basic ML but were unfamiliar with AML. 
We aimed to investigate two aspects of \name{} as an educational tool: \textbf{(A1)} whether \name{} is effective for helping learners understand the concepts and impacts of adversarial attacks, and \textbf{(A2)} whether users enjoy using \name{} for learning.
We did not conduct a comparative study due to existing AML educational tools falling short of such comparison due to their inherent limitations:
\begin{itemize}
\item Adversarial-Playground \cite{norton2017adversarial} is a very simple tool that only provides side-by-side comparisons of natural and adversarial images with classification likelihoods. 
Consequently, this tool includes only a fraction of the functionalities inherent to a single component of \name{}, \ie, the Instance-level Attack Explainer. 
\item Bluff \cite{das2020bluff} only visualizes the internal neuron pathways, which is a divergent focus from the educational scope of \name{}. 
While Bluff is for those interested in the very low-level details of individual neuron behaviors, \name{} is designed to provide a more approachable and practical understanding of the attacks through multiple lenses such as data embeddings, confidence scores, and accuracy degradation.  

\end{itemize}
Since current tools either serve fundamentally different purposes or offer a limited subset of the functionalities provided by \name{}, there currently exists no suitable baseline tool for a meaningful and fair comparison. 
\re{It is also important to note that \name{} is primarily designed to augment, rather than replace, traditional teaching by introducing a layer of interactive teaching. 
As such, by following a similar methodology in \cite{wang2020cnn,kahng2019does}, we opted not to include raw teaching as a baseline, as we believe the unique values brought by \name{} arise from its ability to engage learners in a hands-on manner, which fundamentally differs from the more static, passive learning experience provided by traditional teaching. }

\subsection{Study Setup}

\textbf{Participants and Apparatus. }
We recruited 12 participants (P1 $\sim$ P12; 10 men, two women; aged 21$\sim$31) from a local university. 
They came from different areas of study such as computer science, transportation engineering, and data science. 
All reported having a background in ML but were unfamiliar with AML. 
Specifically, on a 7-point Likert scale (self-rated; 1=``Novice'', 7=``Expert''), we recruited participants that satisfied all the following constraints: ML experience $\geq$ 2, AML experience $\leq$ 2, completion of $\geq$ 1 ML project, completion of $\leq$ 1 AML project.
Their median ML experience was 4 (\iqr{2}), and their median AML experience was 1 (\iqr{0.25}).
The median number of ML projects completed was 2.5 (\iqr{2.25}), while the median number of AML projects completed was 0 (\iqr{0}).
They interacted with \name{} on provided laptops in-person.

\textbf{Task and Procedure. } 
We loaded \name{} with CIFAR-10 testing data perturbed by FGSM in varying degrees to investigate the participants' learning of the properties and impacts of the attack. 
We selected FGSM for our first evaluation study based on recommendations from all interviewed AML instructors/learners, citing it as ideal for introducing AML concepts.
E2 mentioned, \qt{For learners, it is essential to start with foundational methods like FGSM given its straightforward and basic nature.}
S2 agreed that \qt{When teaching learners, it is best to use simpler attacks like FGSM.}
We measured the participants' learning through a pre-quiz before their interaction with \name{}, followed by a post-quiz afterwards to assess the amount of knowledge acquired through the use of our tool.
The quizzes were collaboratively designed with a renowned AML researcher/instructor who co-authored TRADES, the state-of-the-art adversarial training method against evasion attacks that won first place in the robust model track of NeurIPS 2018 Adversarial Vision Challenge \cite{zhang2019theoretically}.
Prior to interacting with \name{}, we asked the participants to complete the pre-quiz that consisted of 9 questions to assess their ML background and knowledge in AML.  
These questions included 4 checker questions on basic ML and 5 questions that would be taught by \name{} \wsicon{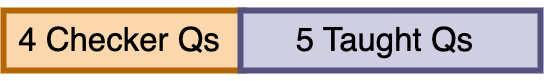}.
The checker questions were to ensure participants' self-reported expertise aligned with their background and to assess their attention during the study. 
After the pre-quiz, we introduced \name{} to the participants and provided them with 5 minutes to go through the beginning tutorial in \name{}.
The tutorial contained basic background on adversarial attacks (\eg, what an adversarial attack is) and guidance on navigating through the different components of \name{}'s interfaces.
Participants were also given the freedom to revisit these tutorials, which are included as part of the General Information Provider, at any point during their interaction with \name{}. 
Following the tutorials, we provided the participants with 30 minutes to interact with \name{} freely. 
We instructed the participants to use \name{} to learn about the FGSM attack as much as they could, and informed them that there would be a follow-up post-quiz to assess how much they had learned.
Next, we asked the participants to complete a 7-point Likert scale post-questionnaire (6 questions), which collected their opinions on the learning and usability aspects of \name{}.
We then asked them to complete the post-quiz (19 questions), which comprised of the 9 original questions from the pre-quiz, along with 10 new questions that were taught by \name{} \wsicon{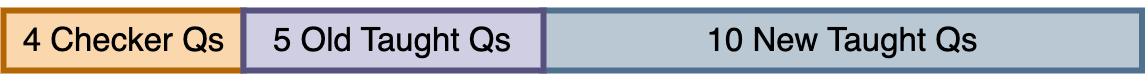}.
We ended the study with a qualitative interview that further asked for their thoughts and opinions on \name{}. 

The user study took about one hour and the participants received $\$$15 for their effort. 
They were informed that the top 3 performers of both the pre-quiz and post-quiz would be awarded an additional $\$$10.

\begin{table*}[t]
\caption{The results of the paired t-tests and the quiz averages of our participants (filtered \& all). Our results show that \name{} has a strong learning effect on both filtered and all participants. ${}^{\ast}$OQ (``old questions''): 9 questions from the pre-quiz that are also included in the post-quiz. ${}^{\dagger}$NQ (``new questions''): 10 questions that are newly added in the post-quiz. ${}^{\ddagger}$Average of total quiz (checkers + taught) scores.}
\label{tab:stats-table}
\vspace{-3mm}
\centering
\setlength{\tabcolsep}{0.5em} %
{\renewcommand{\arraystretch}{0.9}%
\resizebox{\textwidth}{!}{%
\begin{NiceTabular}{c|ccccc|cccc}[cell-space-limits=3pt]
\toprule
\multirow{3.5}{*}{\begin{tabular}[c]{@{}c@{}} \end{tabular}} &
  \multicolumn{5}{c|}{\textbf{Paired T-Tests}} &
  \multicolumn{4}{c}{\textbf{Quiz Averages}} \\ \cmidrule{2-10} 
 &
  \multicolumn{1}{c|}{\begin{tabular}[c]{@{}c@{}}Pre-quiz vs. \\ Post-quiz\end{tabular}} &
  \multicolumn{1}{c|}{\begin{tabular}[c]{@{}c@{}}Pre-quiz vs. \\ Post-quiz OQ$^{\ast}$\end{tabular}} &
  \multicolumn{1}{c|}{\begin{tabular}[c]{@{}c@{}}Pre-quiz vs.\\ Post-quiz NQ$^{\dagger}$\end{tabular}} &
  \multicolumn{1}{c|}{\begin{tabular}[c]{@{}c@{}}Pre-quiz Taught vs.\\ Post-quiz Taught\end{tabular}} &
  \begin{tabular}[c]{@{}c@{}}Pre-quiz Checkers vs.\\ Post-quiz Checkers\end{tabular} &
  \multicolumn{1}{c|}{\begin{tabular}[c]{@{}c@{}}Pre-Quiz \\ Checkers\end{tabular}} &
  \multicolumn{1}{c|}{\begin{tabular}[c]{@{}c@{}}Pre-Quiz \\ Taught\end{tabular}} &
  \multicolumn{1}{c|}{\begin{tabular}[c]{@{}c@{}}Post-Quiz\\ Checkers\end{tabular}} &
  \begin{tabular}[c]{@{}c@{}}Post-Quiz\\ Taught\end{tabular} \\ \midrule
\multirow{2.5}{*}{\begin{tabular}[c]{@{}c@{}}\textbf{Filtered} \\ (10 Participants)\end{tabular}} &
  \multicolumn{1}{c|}{\multirow{2.5}{*}{\begin{tabular}[c]{@{}c@{}}\textit{t} = -5.264, \\ \textit{p} = 0.00052\end{tabular}}} &
  \multicolumn{1}{c|}{\multirow{2.5}{*}{\begin{tabular}[c]{@{}c@{}}\textit{t} = -6.128, \\ \textit{p} = 0.00017\end{tabular}}} &
  \multicolumn{1}{c|}{\multirow{2.5}{*}{\begin{tabular}[c]{@{}c@{}}\textit{t} = -4.229, \\ \textit{p} = 0.00221\end{tabular}}} &
  \multicolumn{1}{c|}{\multirow{2.5}{*}{\begin{tabular}[c]{@{}c@{}}\textit{t} = -6.482, \\ \textit{p} = 0.00011\end{tabular}}} &
  \multirow{2.5}{*}{\begin{tabular}[c]{@{}c@{}}\textit{t} = 1.0, \\ \textit{p} = 0.34344\end{tabular}} &
  \multicolumn{1}{c|}{\begin{tabular}[c]{@{}c@{}}85\% \\ ($\sigma$ = 21.08\%)\end{tabular}} &
  \multicolumn{1}{c|}{\begin{tabular}[c]{@{}c@{}}50\% \\ ($\sigma$ = 17\%)\end{tabular}} &
  \multicolumn{1}{c|}{\begin{tabular}[c]{@{}c@{}}82.5\% \\ ($\sigma$ = 26.48\%)\end{tabular}} &
  \begin{tabular}[c]{@{}c@{}}93.33\% \\ ($\sigma$ = 6.28\%)\end{tabular} \\ \cmidrule{7-10} 
 &
  \multicolumn{1}{c|}{} &
  \multicolumn{1}{c|}{} &
  \multicolumn{1}{c|}{} &
  \multicolumn{1}{c|}{} &
   &
  \multicolumn{2}{c|}{\begin{tabular}[c]{@{}c@{}}65.56\% ($\sigma$ = 16.93\%)$^{\ddagger}$\end{tabular}} &
  \multicolumn{2}{c}{\begin{tabular}[c]{@{}c@{}}91.05\% ($\sigma$ = 7.46\%)$^{\ddagger}$\end{tabular}} \\ \midrule
\multirow{2.5}{*}{\begin{tabular}[c]{@{}c@{}}\textbf{All}\\ (12 Participants)\end{tabular}} &
  \multicolumn{1}{c|}{\multirow{2.5}{*}{\begin{tabular}[c]{@{}c@{}}\textit{t} = -6.225, \\ \textit{p} = 0.00006\end{tabular}}} &
  \multicolumn{1}{c|}{\multirow{2.5}{*}{\begin{tabular}[c]{@{}c@{}}\textit{t} = -6.661, \\ \textit{p} = 0.00004\end{tabular}}} &
  \multicolumn{1}{c|}{\multirow{2.5}{*}{\begin{tabular}[c]{@{}c@{}}\textit{t} = -5.197, \\ \textit{p} = 0.0003\end{tabular}}} &
  \multicolumn{1}{c|}{\multirow{2.5}{*}{\begin{tabular}[c]{@{}c@{}}\textit{t} = -5.88, \\ \textit{p} = 0.00011\end{tabular}}} &
  \multirow{2.5}{*}{\begin{tabular}[c]{@{}c@{}}\textit{t} = -0.561, \\ \textit{p} = 0.5863\end{tabular}} &
  \multicolumn{1}{c|}{\begin{tabular}[c]{@{}c@{}}75\% \\ ($\sigma$ = 30.15\%)\end{tabular}} &
  \multicolumn{1}{c|}{\begin{tabular}[c]{@{}c@{}}51.67\% \\ ($\sigma$ = 15.86\%)\end{tabular}} &
  \multicolumn{1}{c|}{\begin{tabular}[c]{@{}c@{}}77.08\% \\ ($\sigma$ = 27.09\%)\end{tabular}} &
  \begin{tabular}[c]{@{}c@{}}90\% \\ ($\sigma$ = 10.05\%)\end{tabular} \\ \cmidrule{7-10} 
 &
  \multicolumn{1}{c|}{} &
  \multicolumn{1}{c|}{} &
  \multicolumn{1}{c|}{} &
  \multicolumn{1}{c|}{} &
   &
  \multicolumn{2}{c|}{\begin{tabular}[c]{@{}c@{}}62.04\% ($\sigma$ = 17.38\%)$^{\ddagger}$\end{tabular}} &
  \multicolumn{2}{c}{\begin{tabular}[c]{@{}c@{}}87.28\% ($\sigma$ = 11.32\%)$^{\ddagger}$\end{tabular}} \\ 
\bottomrule
\end{NiceTabular}}}
\end{table*}

\subsection{Results and Analysis: Task Performance}

Out of 12 participants, we removed two whose pre-quiz checker scores were below 50\%.
On average, the 10 remaining participants spent 3.97 minutes (\sd{0.07}) on the pre-quiz, 16.17 minutes (\sd{0.21}) on their interaction with \name{}, and 5.17 minutes (\sd{0.10}) on the post-quiz.
Before interacting with \name{}, the participants had an average pre-quiz score of 65.56\% (\sd{16.93\%}), and 50\% (\sd{17\%}) if excluding the checker questions.
After, the participants earned an average post-quiz score of 91.05\% (\sd{7.46\%}), and 93.33\% (\sd{6.28\%}) if excluding the checker questions.
While the difference between the mean pre-quiz and post-quiz scores clearly indicates \name{}'s effectiveness in enabling learning, we further answered \textbf{A1} by performing several paired t-tests on our collected quantitative data. 

Our first paired t-test shows a significant difference between the participants' overall pre-quiz and post-quiz performance (\ttest{-5.264}{0.00052}); the difference is also significant in the second paired t-test when the checker questions are excluded (\ttest{-6.482}{0.00011}).
Both results indicate a strong performance improvement after interaction with \name{}.
A third paired t-test shows a significant difference between their performance on the same 9 questions in the pre-quiz and post-quiz (\ttest{-6.128}{0.00017}).
This indicates that the participants have successfully learned the answers to the questions that were originally included in the pre-quiz.
Similarly, a significant difference can be observed between the participants' pre-quiz performance and their performance on the 10 newly added questions in the post-quiz (\ttest{-4.229}{0.00221}).
This shows that the participants have picked up additional knowledge that was not mentioned in the pre-quiz during their interaction with \name{}. %
Lastly, another paired t-test was performed between their performance on the same checker questions in the pre-quiz and post-quiz and no significant difference was found (\ttest{1.0}{0.34344}).
In conjunction with the fact that all 10 qualified participants scored a minimum of 50\% on the pre-quiz checker questions, this suggests that our participants maintained consistency in their checker responses and did not select answers randomly.

We repeated our statistical tests on all 12 participants, including the two participants who were originally excluded, and our results still demonstrate a strong learning effect (\autoref{tab:stats-table}).
This finding suggests that while \name{} is primarily designed for learners with a basic ML background who are new to AML, it benefits not only the intended users but also proves effective for those without fundamental ML knowledge seeking to understand adversarial attacks.
The ability of \name{} to accommodate a wider audience further emphasizes its value as an educational tool, extending its potential impact by making complex AML concepts more approachable even to those just beginning to explore the field of ML.
The full results of all our paired t-tests and the quiz averages of the participants are shown in \autoref{tab:stats-table}.

\begin{figure}[tb]
\centering
\includegraphics[width=0.9\columnwidth]{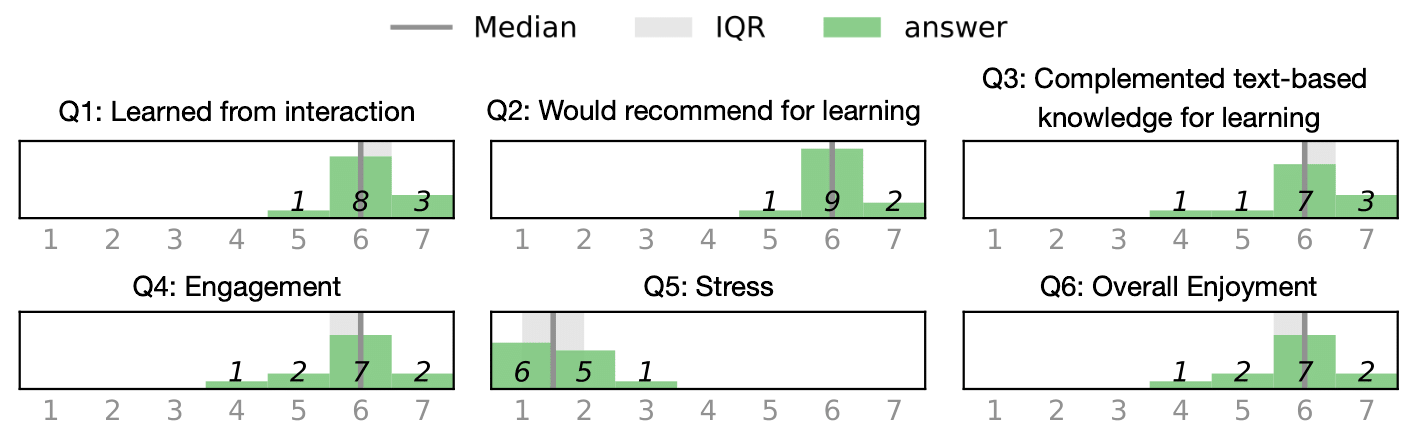}
\caption{Participants' questionnaire ratings (1 = ``strongly disagree''; 7 = ``strongly agree'') on the learning and usability aspects of \name{}. 
}
\label{fig:stats_vis}
\vspace{-5mm}
\end{figure}

\subsection{Results and Analysis: Participants' Feedback}

To further investigate \textbf{A1} and \textbf{A2}, we analyzed the participants' post-questionnaire responses on a 7-point Likert scale (\autoref{fig:stats_vis})
and their qualitative feedback from the interviews on the learning and usability of \name{}. 
The questionnaire asked users whether they: Q1) learned about adversarial attacks through \name{}, Q2) would recommend \name{} for learning, Q3) found \name{} complemented text-based knowledge, Q4) felt engaged, Q5) felt stressed, and Q6) enjoyed overall interaction with \name{}. 

For Q1, all participants agreed that they had learned about adversarial attacks through interacting with \name{} (\md{6}, \iqr{0.5}) and gave a positive rating ($\geq 5$) on \name{}'s learning effect. 
Specifically, the participants felt that \name{} offered comprehensive visualizations and found the explanations very easy to understand.
P3 stated that \qt{\name{} teaches all aspects of adversarial attacks very thoroughly,} and P8 commented that \qt{The clear explanations made the learning process much easier.}
The participants also thought that \name{}'s visualizations were highly informative for them to understand the key attack properties and the underlying attack process.

Similarly, for Q2, all participants stated that they would recommend \name{} to others for learning AML (\md{6}, \iqr{0}). 
They believed that \name{} would be highly beneficial for beginners and can serve as an effective entry point to those interested in learning about adversarial attacks.
P2 thought that \qt{\name{} is a valuable educational tool for illustrating the attacks,} and P5 believed that \qt{\name{} is great for beginners, it can teach them a lot about the attack process.}
To further strengthen \name{} as a learning tool, P5 suggested visualizing the internal attack process in more detail, such as how gradients are modified. 

Q3's ratings show that \name{} complemented the provided text-based knowledge in the General Information Provider well for learning (\md{6}, \iqr{0.5}). 
From this, some participants believed that \name{} could be used in conjunction with text-based documents, such as textbooks, to \pqt{improve learning experience in traditional classroom settings}{P1}.
Other participants felt that \name{} was sufficient on its own for explaining adversarial attacks.
\pqt{I don't think \name{} needs any additional complementary materials. Its visualizations are enough to thoroughly explain the attack logic.}{P4}

Eleven out of 12 participants gave a rating $\geq$ 5 on \name{}'s engagement in Q4 (\md{6}, \iqr{0.5}).
They applauded \name{} for its highly interactive interfaces and enjoyed dynamically experimenting with the perturbation size to see all the real-time changes.
\pqt{It is highly engaging to change the noise level and observe how the resulting image differs.}{P1}
Similarly, P5 stated, \qt{It is fun to see all the points move around in the Data Projectors as I adjust the slider.}
However, one participant, P12, rated \name{}'s engagement a 4 and explained, \qt{In general, the application is good. But as a programmer, I feel like I should be able to get more involved and write custom code directly.}

For Q5, all participants agreed that it was not stressful to interact with \name{} (\md{1.5}, \iqr{1}).
This was likely because \name{} had an interactive tutorial system that provided guidance on \name{}'s functionalities, along with the General Information Provider that offered further assistance.
Moreover, everything \name{} visualized (\eg, 2-D latent space, confidence scores) were familiar to learners who knew ML, thus making \name{} intuitive to learn with.
\pqt{Using \name{} is very simple, I didn't encounter any difficulties. The visualizations are all quite straightforward and intuitive.}{P7}

In general, the participants rated \name{}'s enjoyment positively in Q6 (\md{6}, \iqr{0.5}).
They offered different reasons for why they enjoyed \name{}. 
P9 and P10 claimed that \name{}'s visually appealing interfaces and animations made their interactions entertaining.
P3, P8 \& P10 emphasized the amount of knowledge they gained from \name{} and found the learning experience fruitful.
P4, P6 \& P11 applauded \name{} for its high level of interactivity.
\pqt{I enjoy \name{} because I can do a lot with it. I can investigate different examples, try out different noise levels, and observe how the embedding distribution changes.}{P4}

\section{Interview Study with Experienced Experts/Teachers}

To collect more in-depth qualitative feedback on \name{}, we conducted an interview study with AML experts/teachers, who possess profound knowledge of the key aspects and requirements for understanding adversarial attacks. 
These interviews provided additional insights into how \name{} can be utilized in an educational setting.

\subsection{Study Setup}
In this study, AML experts/teachers were prompted to use \name{} to explore one white-box attack and one black-box attack, FGSM and ZOO, on four different models (VGG-16, VGG-19, standard ResNet-34, \& adversarially trained ResNet-34) with the CIFAR-10 data in a free-form analysis session.
We recruited seven AML experts (E1, E2, E4 $\sim$ E8; all men), six of whom have teaching experience that spans from leading AML seminars to teaching ML courses with AML components.
Each study session began with a 5-minute introduction to the study background and \name{}'s key features.
Next, we presented a task scenario where participants were asked to use \name{} to explore and understand \qt{how the FGSM and ZOO attacks alter the input images to affect the models' performance,} and \qt{how models display varying robustness against the attacks.}
Participants had 30 minutes to explore each attack, and a task list was provided to guide their interaction. 
They were also informed that they could explore the tool freely without following these tasks as long as insights were gathered.
We employed the think-aloud protocol, requiring participants to provide feedback from both the perspectives of \textit{experts/teachers} and \textit{learners}.
An experimenter was responsible for providing help and answering questions regarding the interface, who also observed the experts' interactions and took notes.
After the interaction, a semi-structured interview ($\approx$30 minutes) was conducted to %
gain a better understanding of the participants' thoughts on \name{} in light of the think-aloud feedback and observation gathered previously. 
The participants were compensated $\$$20/hour for the study.

\subsection{Results and Analysis}
All seven experts successfully used \name{} to gain insights into the attacks and expressed a positive sentiment toward it. 
We conducted a thematic analysis on the unstructured feedback gathered during the free-form analysis and the qualitative data provided to us during the semi-structured interviews.
We came up with five systematic themes aligned with our design goals and an additional theme focused on usability, and adopted a deductive approach to identify patterns of them in our data.

\textbf{Visualizations of attack impacts.}
All experts agreed that \name{} can help learners quickly grasp the attack impacts. 
E4 and E6 liked the Robustness Analyzers for illustrating \qt{the overall trend of accuracy changes.} 
E1, E6, and E8 valued the Data Projectors for allowing learners to \qt{see how embeddings are drifted from their original positions.}
They also found the lower-level visualizations highly useful.
E7 noted, \qt{The confidence score view can show learners that ZOO [...] pushes instances just past the decision boundary.}
This confirms that \name{} effectively visualizes the attack impact at multiple levels (\textbf{G1}).
A noted limitation is the occasional difficulty in distinguishing between points misclassified before and due to the attack.
E1 suggested displaying both the original and current predictions in the attack explainer.

\textbf{Generalizability.}
The experts praised \name{} for its generalizability to different attacks and image classifiers. %
E2 explained, \qt{\name{}'s ability to adapt to different attacks and models is vital for learners to truly understand the risks by evaluating against diverse techniques.} 
This confirms that it can effectively help learners assess the variability of models and attack methods (\textbf{G2}). 
Moreover, the experts highlighted that such design simplifies the exploration by providing a more accessible way to investigate different attacks. 
Both E1 and E2 pointed out how \name{} saves learners' time by eliminating the need to code from scratch when exploring different attack strategies on their own models.

\textbf{Evaluation of model robustness.}
The experts believed that \name{} can help learners easily discern their models' strengths and weaknesses.
The model comparison feature was frequently highlighted, with E7 noting that it can reveal that \qt{deeper models do not necessarily excel under attacks.}
Similarly, E5 and E6 commented that it shows that a robust model has \qt{embeddings that barely differ under standard or adversarial conditions.}
These comments affirm \name{}'s capability for detailed visual analysis and model comparison (\textbf{G3}). 
While the experts valued how comprehensive the model visualizations are, E1 and E4 suggested adding comparison of the same model under attacks with different perturbation sizes side by side.

\textbf{Dynamic experimentation with real-time changes.} 
The experts enjoyed dynamically experimenting with the perturbation size and found the real-time visual feedback valuable.
E2 commented, \qt{\name{} answers questions that papers and tutorials may not cover, such as the effects of varying perturbation sizes on model embeddings.}
Furthermore, they believed the integration of dynamic perturbation adjustment and real-time visual feedback offers an engaging learning experience.
E2 explained that learners could play with \name{} for self-learning, while E1 thought teachers could use the tool to \qt{demonstrate attacks in a fun, interactive, and engaging way.}
These observations suggest that \name{} provides a highly interactive learning experience with its perturbation experimentation and real-time feedback (\textbf{G4}).

\textbf{Overall benefits as an educational tool for learners.}
All experts agreed that \name{} is highly beneficial as an educational tool.
E6 stated, \qt{\name{} bridges theory and practice, enhancing learners' understanding [...] and encouraging them to further explore the field.}
They also thought the step-by-step execution would be very informative for AML learners, confirming AdvEx's capability to enable detailed learning of the attack process (\textbf{G5}).
In addition, the experts believed that \name{}'s interfaces would make the learning experience highly enjoyable.
E1 commented, \qt{\name{}'s game-like experience makes learning and evaluating models much easier for learners without too many tedious formulas.}

\textbf{Usability \& beginner-friendly design.}
All experts thought \name{} was very intuitive to pick up.
E5 liked how the beginning tutorial highlighted specific areas of the interface, which helped him easily understand the purpose of each component.
E1 thought learners less experienced with AML could also pick up \name{} easily. 
This confirms that \name{} was successfully integrated with a beginner-friendly design.
The experts also thought that \name{} was very accessible. %
E5 highlighted the zoomable binned aggregation feature and commented, \qt{This feature effectively accommodates different users' available computational power and enable smooth exploration of large-scale data for everyone.}

\section{Discussion}

Here, we discuss the limitations of our current system and outline future directions to enhance our work.
We also present the potential avenues for extending and generalizing our proposed design.

\subsection{Limitations and Future Work}
While our study confirms that \name{} is highly effective in helping learners understand adversarial attacks, it still has several limitations.
Firstly, as commented by our participants, the current Data Projectors (\autoref{fig:teaser}c) allow comparisons of two different models under the same perturbation level, but do not support comparing the same models side by side at different perturbation levels.
Future extensions should enable this type of comparison without adjusting the slider back and forth.
A simple solution is to add additional toggles to the projectors for switching between the different comparison modes.

Secondly, when the perturbation size exceeds 0, distinguishing instances misclassified before the attack from those misclassified due to the attack becomes less intuitive. 
This can be easily mitigated by implementing additional visual encodings such as using different shapes (triangles and crosses) to represent the two types of misclassifications. 
However, this approach may increase the cognitive load of users. %
But an optional filtering feature can be added to allow users to focus only on either type of misclassification.

\re{Thirdly, due to color distinguishability \cite{munzner2014visualization}, \name{} is currently limited to handling datasets with $\le$ 12 classes or subsets of larger datasets. 
While our tool could theoretically support more colors, increasing the number beyond twelve would reduce the effectiveness of exploration due to colors appearing too similar. 
Nonetheless, Our approach aligns with common encoding methods used in existing visualization tools \cite{ren2016squares, kahng2017cti}, and \name{} accommodates datasets with a larger number of classes by allowing learners to focus on exploring a smaller subset of more critical classes. 
In future work, alternative encoding methods (\eg, using a combination of color with other visual cues like shapes or patterns) could be explored to potentially expand the number of distinguishable classes while preserving  the clarity and usability of the visualization. }

Finally, the evaluation of \name{} can be further enhanced.
A larger sample size should be obtained to better evaluate \name{}'s effectiveness.
Also, the current study was designed with a few selected models, and only the FGSM and ZOO attacks with the CIFAR-10 data were used to assess the learning effect and usability of \name{}.
In the future, deployment studies with other types of attacks and datasets should also be conducted to investigate how \name{} can be used in various real-world domains.
This will thoroughly examine the strengths and weaknesses of \name{}, and help us understand how \name{} can be potentially incorporated into learners' model development workflows.

\subsection{Generalization and Extension}

\textbf{Generalization to other applications}. 
We designed \name{} as a system for visualizing adversarial attacks, but the tool is flexible enough to be adapted to visualize other data augmentations in image classification. 
For example, \name{} can be extended to visualize noise applications (\eg, Gaussian noise, salt-and-pepper noise) or other image degradations (\eg, motion blur, JPEG compression).
Learners may use \name{} to understand how visual quality impacts the performance of models in those scenarios.
\re{While our focus is primarily adversarial attacks in image classification, we acknowledge that there are similar attacks in other prevalent ML tasks, such as object detection \cite{xie2017adversarial}, audio processing \cite{carlini2018audio}, and NLP \cite{zhang2020adversarial}. 
Although visualizing attacks in these domains is outside our current scope, we believe there is potential for our approach to be generalized to these domains. 
Components like Robustness Analyzers, Data Projectors, and confidence score view are already well-generalized and provide an intuitive way to explore and assess models' accuracy, feature representations, and output probabilities across different domains, making such designs easily extendable to other ML tasks. 
Learners can use these views to observe embedding movements and output changes when subtle background noises are applied to audio inputs, or after letters in text are maliciously replaced, deleted, or swapped. 
For other components like Instance-level Attack Explainer, future enhancement may be made to generalize them to different types of data instances. 
For example, further implementations could be made to detect input data type and adjust the layout of the attack explainer accordingly to better accommodate the data.
For example, for NLP, the view can be adjusted to highlight which letters in the text instances have been modified. 
For object detection, the view can display the before and after of bounding box manipulation that causes models to miss the targets.
The extension of our tool to these areas could further heighten its educational value and provide additional knowledge to learners into adversarial attacks across a spectrum of ML applications. }

\textbf{Extension to other ML concepts.} 
\name{} leverages a balanced combination of active visualizations and passive text-based information to help learners understand AML, and this design can be applied to visualization tools for learning other ML concepts.
In fact, many existing tools (\eg, \cite{norton2017adversarial,das2020bluff}) only focus on their interactive visualizations and place little emphasis on text-based information, not providing enough guidance and background knowledge to the users.
On the other hand, interactive articles \cite{hohman2020communicating} usually involve mainly text and provide insufficient visualizations.   
\name{} places more balanced weights on both components, ensuring that the users may gain detailed and accurate AML knowledge from our General Information Provider (\autoref{fig:teaser}e) in addition to exploration with the visualizations.
Our design not only reinforces learning by presenting content in multiple formats, but also allows the learners to quickly grasp complex topics requiring visual interpretations, which could shed light on future research on the spectrum of modalities for teaching ML concepts.

\vspace{-2mm}

\section{Conclusion}

We have presented \name{}, an interactive visualization tool for novice AML learners to explore and understand adversarial attacks.
Based on the design guidelines derived from user interviews, we designed \name{} to provide learners with detailed attack visualizations at multiple levels, highlighting attack's properties and effects on different image classifiers. 
Our design addresses the limitations of existing tools, which lack comprehensiveness and generalizability when visualizing the attacks.
We quantitatively and qualitatively assessed \name{} in a two-part evaluation, including a user study with 12 AML learners and an interview study with seven AML experts/teachers.
Our results indicate that \name{} is not only highly effective as an educational tool, but also provides an engaging and enjoyable learning experience, thus highlighting its overall benefits for AML learners.
Additionally, we discuss the future directions to enhance our work and present potential avenues to extend and generalize \name{} to other applications.

\bibliographystyle{ACM-Reference-Format}
\bibliography{main_bib}

\end{document}